# Low-Rank Mechanism: Optimizing Batch Queries under Differential Privacy

Ganzhao Yuan[1]  Zhenjie Zhang[2]  Marianne Winslett[2,3]  Xiaokui Xiao[4]  Yin Yang[2]  Zhifeng Hao[1]

[1]School of Computer Science & Engineering
South China University of Technology
{y.ganzhao,mazfhao}@scut.edu.cn

[2]Advanced Digital Sciences Center
Illinois at Singapore Pte. Ltd.
{zhenjie,yin.yang}@adsc.com.sg

[3]Dept. of Computer Science
University of Illinois at Urbana-Champaign
winslett@illinois.edu

[4]School of Computer Engineering
Nanyang Technological University
xkxiao@ntu.edu.sg

## ABSTRACT

Differential privacy is a promising privacy-preserving paradigm for statistical query processing over sensitive data. It works by injecting random noise into each query result, such that it is provably hard for the adversary to infer the presence or absence of any individual record from the published noisy results. The main objective in differentially private query processing is to maximize the accuracy of the query results, while satisfying the privacy guarantees. Previous work, notably the matrix mechanism [16], has suggested that processing a batch of correlated queries as a whole can potentially achieve considerable accuracy gains, compared to answering them individually. However, as we point out in this paper, the matrix mechanism is mainly of theoretical interest; in particular, several inherent problems in its design limit its accuracy in practice, which almost never exceeds that of naïve methods. In fact, we are not aware of any existing solution that can effectively optimize a query batch under differential privacy. Motivated by this, we propose the *Low-Rank Mechanism* (LRM), the first practical differentially private technique for answering batch queries with high accuracy, based on a *low rank approximation* of the workload matrix. We prove that the accuracy provided by LRM is close to the theoretical lower bound for any mechanism to answer a batch of queries under differential privacy. Extensive experiments using real data demonstrate that LRM consistently outperforms state-of-the-art query processing solutions under differential privacy, by large margins.

## 1. INTRODUCTION

Differential privacy [11] is an emerging paradigm for publishing statistical information over sensitive data, with strong and rigorous guarantees on individuals' privacy. Since its proposal, differential privacy has attracted extensive research efforts, such as cryptography [11], algorithms [12, 14, 21], databases [8, 15, 16, 24, 27, 28, 29], data mining [1, 13] and machine learning [3, 4, 25]. The main idea of differential privacy is to inject random noise into aggregate query results, such that the adversary cannot infer, with high confidence, the presence or absence of any given record $r$ in the dataset, even if the adversary knows all other records in the dataset except for $r$. This paper follows a popular definition of differential privacy, called $\epsilon$-differential privacy, in which the adversary's maximum confidence in inferring private information is controlled by a user-specified parameter $\epsilon$ called the *privacy budget*. Given $\epsilon$, the main goal of query processing under $\epsilon$-differential privacy is to maximize the utility/accuracy of the (noisy) query answers, while satisfying the above privacy requirements.

This work focuses on a common class of queries called *linear counting queries*, which is the basic operation in many statistical analyses. Similar ideas apply to other types of linear queries, e.g., linear sums. Figure 1(a) illustrates an example electronic medical record database, where each record corresponds to an individual. Figure 1(b) shows the exact number of HIV+ patients in each state, which we refer to as *unit counts*. A linear counting query in this example can be any linear combination of the unit counts. For instance, let $x_{NY}$, $x_{NJ}$, $x_{CA}$, $x_{WA}$ be the patient counts in states NY, NJ, CA, and WA respectively; one possible linear counting query is $x_{NY} + x_{NJ} + x_{CA} + x_{WA}$, which computes the total number of HIV+ patients in the four states listed in our example. Another example linear counting query is $x_{NY}/19 + x_{NJ}/8 + x_{CA}/37$, which calculates the weighted average of patient counts in states NY, NJ and CA, with weights set according to their respective population sizes. In general, we are given a database with $n$ unit counts, and a batch $QS$ of $m$ linear counting queries. The goal is to answer all queries in $QS$ under $\epsilon$-differential privacy, and maximize the expected overall accuracy of the queries.

| Name  | State | HIV+ |
|-------|-------|------|
| Alice | NY    | Yes  |
| Bob   | NJ    | Yes  |
| Carol | NY    | Yes  |
| Dave  | CA    | Yes  |
| ...   |       |      |

| State | # of HIV+ patients |
|-------|--------------------|
| NY    | 82,700             |
| NJ    | 19,000             |
| CA    | 67,000             |
| WA    | 5,900              |
| ...   |                    |

(a) Patient records  (b) Statistics on HIV+ patients

**Figure 1: Example medical record database**

Straightforward approaches to answering a batch of linear counting queries usually lead to sub-optimal result accuracy. One naïve solution, referred to as *noise on queries* (*NOQ*), is to process each query independently, e.g., using the Laplace Mechanism [11]. This method fails to exploit the *correlations* between different queries. Consider a batch of three different queries $q_1 = x_{NY} + x_{NJ} + x_{CA} + x_{WA}$, $q_2 = x_{NY} + x_{NJ}$, $q_3 = x_{CA} + x_{WA}$. Clearly, the three queries are correlated since $q_1 = q_2 + q_3$. Thus, an alternative strategy for answering these queries is to process only $q_2$ and $q_3$, and use their sum to answer $q_1$. As will be explained in Section 3, the amount of noise added to query results depends upon the *sensitivity* of the query set, which is defined as the maximum possible total change in query results caused by adding or removing a single record in the original database. In our example, the sensitivity of the query set $\{q_2, q_3\}$ is 1, because adding/removing a patient record in Figure 1a affects at most one of $q_2$ and $q_3$ (i.e., $q_2$ if the record is associated with state NY or NJ, and $q_3$ if the state is CA or WA), by exactly 1. On the other hand, the query set $\{q_1, q_2, q_3\}$ has a sensitivity of 2, since a record in the above 4 states affects both $q_1$ and one of $q_2$ and $q_3$. According to the Laplace mechanism, the variance of the added noise to each query is $2\Delta^2/\epsilon^2$, where $\Delta$ is the sensitivity of the query set, and $\epsilon$ is the user-specified privacy budget. Therefore, processing $\{q_1, q_2, q_3\}$ directly incurs a noise variance of $8/\epsilon^2$ for each query; on the other hand, executing $\{q_2, q_3\}$ leads to noise variance of $2/\epsilon^2$ for each of $q_2$ and $q_3$, and their sum $q_1 = q_2 + q_3$ has a noise variance of $2 \times 2/\epsilon^2 = 4/\epsilon^2$. Clearly, the latter method obtains higher accuracy for all queries.

Another simple solution, referred to as *noise on data* (*NOD*), is to process each unit count under differential privacy, and combine them to answer the given linear counting queries. Continuing the example, this method computes the noisy counts for $x_{NY}$, $x_{NJ}$, $x_{CA}$ and $x_{WA}$, and uses their linear combinations to answer $q_1$, $q_2$, and $q_3$. This approach overlooks the correlations between different unit counts. In our example, $x_{NY}$ and $x_{NJ}$ (and similarly, $x_{CA}$ and $x_{WA}$) are either both present or both absent in every query, and, thus, can be seen as a single entity. Processing them as independent queries incurs unnecessary accuracy costs when re-combining them. In the example, NOD adds noise with variance $2/\epsilon^2$ to each unit count, and their combinations to answer $q_1$, $q_2$, and $q_3$ have noise variance $8/\epsilon^2$, $4/\epsilon^2$ and $4/\epsilon^2$ respectively. NOD's result utility is also worse than the above-mentioned strategy of processing $q_2$ and $q_3$, and adding their results to answer $q_1$.

In general, the query set $QS$ may exhibit complex correlations among different queries and among different unit counts. As a consequence, it is non-trivial to obtain the best strategy to answer $QS$ under differential privacy. For instance, consider the following query set:

$$\begin{aligned} q_1 &= 2x_{NJ} + x_{CA} + x_{WA} \\ q_2 &= x_{NJ} + 2x_{WA} \\ q_3 &= x_{NY} + 2x_{CA} + 2x_{WA} \end{aligned}$$

NOQ is clearly a poor choice, since it incurs a sensitivity of 5 (e.g., a record of state WA affects $q_1$ by 1, and $q_2$ and $q_3$ by 2 each). The sensitivity of NOD remains 1, and it answers $q_1$, $q_2$, and $q_3$ with noise variance $12/\epsilon^2$, $10/\epsilon^2$ and $18/\epsilon^2$ respectively, leading to a sum-square error (SSE) of $40/\epsilon^2$. The optimal strategy in terms of SSE in this case computes the noisy results of $x_{NJ}$ and $x_{WA}$, as well as $q'_1 = x_{NY}/3 + x_{CA}$, and $q'_2 = 2x_{NY}/3$. Then, it obtains the results for $q_1$, $q_2$, and $q_3$ as follows.

$$\begin{aligned} q_1 &= q'_1 + 2x_{NJ} + x_{WA} - q'_2/2 \\ q_2 &= x_{NJ} + 2x_{WA} \\ q_3 &= 2q'_1 + 2x_{WA} + q'_2/2 \end{aligned}$$

The sensitivity of the above method is also 1, and it answers $q_1$, $q_2$, and $q_3$ with noise variance $12.5/\epsilon^2$, $10/\epsilon^2$ and $16.5/\epsilon^2$ respectively, resulting an SSE of $39/\epsilon^2$. Observe that the there is no simple pattern in the query set or the optimal strategy. Since there is an infinite space of possible strategies, searching for the best one is a challenging problem.

Li et al. [16] investigate the problem of identifying a good strategy for answering these kinds of query sets under differential privacy, and they propose a solution called the *matrix mechanism*. This solution, however, is mainly of theoretical interests due to the two reasons. First, it incurs an enormous computational overhead that limits its applicability to very small data and query sets. Second and more importantly, the matrix mechanism often generates sub-optimal strategies for answering queries; in particular, its practical performance (in terms of query accuracy) almost never exceeds that of the naïve solution NOD. Motivated by this, we propose a novel solution called the *low-rank mechanism* (*LRM*), based on the theory of low-rank matrix approximation. We prove that the accuracy provided by LRM is within a constant factor of the theoretical lower bound established in [14]. Extensive experiments demonstrate that LRM significantly outperforms existing solutions in terms of result accuracy, sometimes by orders of magnitude.

The rest of the paper is organized as follows. Section 2 reviews previous studies on differential privacy. Section 3 provides formal definitions for our problem. Section 4 presents the mechanism formulation of LRM, and analyzes its optimality. Section 5 discusses how to solve the optimization problem in LRM. Section 6 verifies the superiority of our proposal through an extensive experimental study. Finally, Section 7 concludes the paper.

## 2. RELATED WORK

Section 2.1 surveys general purpose mechanisms for enforcing differential privacy. Section 2.2 presents our main competitor, the matrix mechanism [16].

### 2.1 Differential Privacy Mechanisms

Differential privacy was first formally presented in [11], though some previous studies have informally used similar models, e.g., [9]. The Laplace mechanism [11] is the first generic mechanism for enforcing differential privacy, which works when the output domain is a multi-dimensional Euclidean space. McSherry and Talwar [21] propose the exponential mechanism, which applies to any problem with a measurable output space. The generality of the exponential mechanism makes it an important tool in the design of many other differentially private algorithms, e.g., [6, 29, 21].

Linear query processing is of particular interest in both the theory and database communities, due to its wide range of applications. To minimize the error of linear queries under differential privacy requirements, several methods try to build a synopsis of the original database, such as Fourier transformations [24], wavelets [28] and hierarchical trees [15]. By publishing a noisy synopsis under $\epsilon$-differential privacy, these methods are capable of answering an arbitrary number of linear queries. However, most of these methods obtain good accuracy only when the query selection criterion is a continuous range; meanwhile, since these methods are not workload-aware, their performance for a specific workload tends to be sub-optimal.

The compressive mechanism [17] reduces the amount of noise necessary to satisfy differential privacy, by utilizing the sparsity of the dataset under certain transformations. The main idea is to use a technique called compressive sensing to compress a sparse representation of the data into a compact synopsis, and inject noise into the much smaller synopsis instead of the original data. After that, the method reconstructs the original data by applying the decoding algorithm of compressive sensing to the noisy synopsis. The

1353

result provides significantly higher utility, while satisfying differential privacy requirements.

Several theoretical studies have derived lower bounds for the noise level for processing linear queries under differential privacy. Notably, Dinur and Nissim [9] prove that any perturbation mechanism with maximal noise of scale $o(n)$ cannot possibly preserve personal privacy, if the adversary is allowed to ask all possible linear queries, and has exponential computation capacity. By reducing the computation capacity of the adversary to polynomial-bounded Turing machines, they show that an error scale $\Omega(\sqrt{n})$ is necessary to protect any individual' privacy.

More recently, Hardt and Talwar [14] have significantly tightened the error lower bound for answering a batch of linear queries under differential privacy. Given a batch of $m$ linear queries, they prove that any $\epsilon$-differential privacy mechanism leads to squared error of at least $\Omega(\epsilon^{-2}m^3Vol(W))$, where $Vol(W)$ is the volume of the convex body obtained by transforming the $\mathcal{L}_1$-unit ball into $m$-dimensional space using the linear transformations in the workload $W$. They also propose a mechanism for differential privacy whose error level almost reaches this lower bound. However, their mechanism relies on uniform sampling in a high-dimensional convex body, which, although it theoretically takes polynomial time, is too expensive to be of practical use. This paper extends their analysis to low-rank workload matrices.

Besides linear queries, differential privacy is also applicable to more complex queries in various research areas, due to its strong privacy guarantee. In the field of data mining, Friedman and Schuster [13] propose the first algorithm for building a decision tree under differential privacy. Mohammed et al. [22] study the same problem, and propose an improved solution based on a generalization strategy coupled with the exponential mechanism. Ding et al. [8] investigate the problem of differentially private data cube publication. They present a randomized materialized view selection algorithm, which reduces the overall error, and preserves data consistency.

In the database literature, a plethora of methods have been proposed to optimize the accuracy of differentially private query processing. Cormode et al. [6] investigate the problem of multi-dimensional indexing under differential privacy, with the novel idea of assigning different amounts of privacy budget to different levels of the index. Xu et al. [29] optimize the procedure of building a differentially private histogram, with an interesting combination of a dynamic programming algorithm for optimal histogram computation and the exponential mechanism.

Differential privacy is also becoming a hot topic in the machine learning community, especially for learning tasks involving sensitive information, e.g., medical records. In [4], Chaudhuri et al. propose a generic differentially private learning algorithm, which requires strong convexity of the objective function. Rubinstein et al. [25] study the problem of SVM learning on sensitive data, and propose an algorithm to perturb the kernel matrix with performance guarantees, when the loss function satisfies the $l$-Lipschitz continuity property. General differential privacy techniques have also been applied to real systems, such as network trace analysis [19] and private recommender systems [20].

## 2.2 Matrix Mechanism

Li et al. [16] propose the matrix mechanism, which targets the same problem as this work, i.e., answering a batch of linear queries under differential privacy. Given a workload of linear queries, the matrix mechanism first constructs a *workload matrix* $W$ of size $m \times n$, where $m$ is the number of queries, and $n$ is the number of unit counts. The construction of the workload matrix is elaborated further in Section 3. After that, the mechanism searches for a *strategy matrix* $A$ of size $r \times n$, where $r$ is a positive integer. Intuitively, $A$ corresponds to another set of linear queries, such that every query in $W$ can be expressed as a linear combination of the queries in $A$. The matrix mechanism then answers the queries in $A$ under differential privacy, and subsequently uses their noisy results to answer queries in $W$.

The main challenge faced by the matrix mechanism is to identify the strategy matrix $A$ that answers $W$ with the highest accuracy. The solution in [16] is limited to the case where (i) $A$ has a pseudo-inverse $A^\dagger$; and (ii) $A$ is optimized based on the $\mathcal{L}_2$ approximation of the objective function. However, one necessary condition for a matrix $A$ to have a pseudo-inverse is that there must be at least as many rows as columns, i.e., $r \geq n$. This requirement seriously limits the search space for $A$. For instance, imagine an application, akin to that shown in Figure 1, where there are 50 unit counts, each corresponding to a state in the US. Then, the strategy matrix must have at least 50 queries, regardless of how many queries there are in the original workload $W$. None of the strategies used in the example of Figure 1 can be identified by the matrix mechanism, simply because they do not contain enough queries. Furthermore, the optimal answer computed using the modified objective function (i.e., its $\mathcal{L}_2$ approximation) does not necessarily lead to low error according to the original objective function. In fact, throughout our experimental evaluations, we have never found a single setting where the matrix mechanism obtains lower overall error than the naïve solution of injecting noise directly into the unit counts. In addition, the matrix mechanism also incurs a high computational overhead. Overall, the matrix mechanism is mainly of theoretical interest.

## 3. PRELIMINARIES

In this paper, we assume there are $n$ records in a database $D$, i.e., $D = \{x_1, x_2, \ldots, x_n\}$. Each $x_i$ in $D$ is a real number. To facilitate matrix manipulations, in the rest of the paper we use a vector of size $n \times 1$ to denote the database, i.e. $\{x_1, x_2, \ldots, x_n\}^T$. In Figure 1, for example, each record contains the number of HIV+ patients in a state of the USA. A query set $Q$ of cardinality $m$ is a mapping from the database domain to real numbers, i.e., $Q : \mathbb{D} \mapsto \mathbb{R}^m$.

### 3.1 Differential Privacy

A query processing mechanism $M$ is a randomized mapping from $\mathbb{D} \times \mathbb{Q}$ to $\mathbb{R}^m$. Given an arbitrary query set $Q \in \mathbb{Q}$ and a database $D \in \mathbb{D}$, the mechanism $M$ returns a distribution on the query output domain $\mathbb{R}^m$. Two databases $D_1$ and $D_2$ are neighbor databases *iff* they differ on exactly one record, i.e., $D_1 = \{x_1, x_2, \ldots, x_i, \ldots, x_n\}$ and $D_2 = \{x_1, x_2, \ldots, x_i', \ldots, x_n\}$. A randomized mechanism $M$ satisfies $\epsilon$-differential privacy if for every pair of neighbor databases $D_1$ and $D_2$, we have

$$\forall Q \forall R: \Pr(M(Q, D_1) = R) \leq e^\epsilon \Pr(M(Q, D_2) = R) \quad (1)$$

The above inequality implies that the mechanism $M$ always returns similar results on neighbor databases. This limits the adversary's confidence in inferring any record from the output of $M$, even when he or she knows all remaining records in the database.

In [11], Dwork et al. presented a general protocol to implement $\epsilon$-differential privacy, utilizing the concept of *sensitivity*. Given a query set $Q \in \mathbb{Q}$, the sensitivity $\Delta$ is the maximal $\mathcal{L}_1$ distance between the exact query results on any neighbor databases $D_1$ and $D_2$, i.e.

$$\Delta = \max_{D_1, D_2} \|Q(D_1), Q(D_2)\|_1 \quad (2)$$

We emphasize that $\Delta$ only depends on the data domain $\mathbb{D}$ and the query set $Q$, not the actual data. Therefore, we simply assume such a constant $\Delta$ is public knowledge to everyone, including the adversary. The *Laplace Mechanism* [11], $M_L$, outputs a randomized result $R$ on database $D$, following a Laplace distribution with mean $Q(D)$ and magnitude $\frac{\Delta}{\epsilon}$, i.e.,

$$\Pr(M_L(Q, D) = R) \propto \exp\left(\frac{\epsilon}{\Delta}\|R - Q(D)\|_1\right) \quad (3)$$

This is equivalent to adding $m$-dimensional independent Laplace noise, as $Q(D) + Lap\left(\frac{\Delta}{\epsilon}\right)^m$, in which $Lap\left(\frac{\Delta}{\epsilon}\right)$ is a random variable following a zero-mean Laplace distribution with scale $\frac{\Delta}{\epsilon}$. Based on the definition of the Laplace mechanism, the expected squared error of the randomized query answer is $\frac{2m\Delta^2}{\epsilon^2}$, since the variance of $Lap(s)$ is $2s^2$ for any scale $s$. Note that the amount of error only depends on the sensitivity of the queries, regardless of the records in database $D$.

## 3.2 Batch Linear Queries

As mentioned in the introduction, we focus on non-interactive linear queries in this paper. A linear query $q(D)$ is in the form of a linear function over the records in the database. Given a weight vector $\{w_1, w_2, \ldots, w_n\}^T$ of size $n$, the linear query returns the dot product between the weight vector and database vector, i.e.,

$$q(D) = w_1 x_1 + w_2 x_2 + \ldots + w_n x_n$$

We assume a batch of $m$ linear queries, $Q = \{q_1, q_2, \ldots, q_m\}$, is submitted to the database at the same time. The query set $Q$ is thus represented by a *workload matrix* $W$ with $m$ rows and $n$ columns. Each entry $W_{ij}$ in $W$ is the $j$-th coefficient for query $q_i$ on record $x_j$. Using the vector representation of the database, i.e. $D = (x_1, x_2, \ldots, x_n)^T$, the query batch $Q$ can be exactly answered by calculating:

$$Q(D) = WD = \left(\sum_j W_{1j}x_j, \ldots, \sum_j W_{mj}x_j\right)^T$$

Based on the Laplace mechanism, two baseline solutions to enforce $\epsilon$-differential privacy on a query batch with workload $W$ are as follows.

**Noise on data:** This solution, denoted as $M_D$, adds noise to the original data. Given database $D$, $M_D$ generates a noisy database $D'$ using the Laplace mechanism, i.e., $D' = D + Lap\left(\frac{\Delta}{\epsilon}\right)^n$. The query batch $Q$ is then answered by replacing $D$ with $D'$. The whole mechanism can be written in the form of manipulation on random variables, as follows.

$$M_D(Q, D) = WD' = W\left(D + Lap\left(\frac{\Delta}{\epsilon}\right)^n\right) \quad (4)$$

Based on the linearity of expectation, it is straightforward to calculate the expected squared error on the output, $\frac{2\Delta^2}{\epsilon^2}\sum_{i,j} W_{ij}^2$, which is proportional to the squared sum of the entries in $W$.

**Noise on results:** This baseline solution, denoted as $M_R$, adds noise to the query results instead of the original data. Since the queries are linear queries, the sensitivity of the query set is $\Delta' = \max_j \sum_i |W_{ij}|\Delta$, i.e., the highest column absolute sum [16]. Thus, $M_R$ outputs the following random results.

$$M_R(Q, D) = WD + Lap\left(\frac{\Delta'}{\epsilon}\right)^m \quad (5)$$

Similarly, the expected squared error of the mechanism on query $Q$ is $2m\Delta'^2\epsilon^{-2} = 2m\max_j \sum_i W_{ij}^2 \Delta^2 \epsilon^{-2}$. By comparing their expected squared errors, we derive that $M_R$ outperforms $M_D$ by expectation, *iff* $m\max_j \sum_i W_{ij}^2 < \sum_j \sum_i W_{ij}^2$. When $m \geq n$, this inequality can never hold, implying that $M_R$ is more effective only when $m$ is smaller than $n$.

## 3.3 Low Rank Matrices

For any square matrix $A = \{A_{ij}\}$ of size $n \times n$, the trace of the matrix is the sum of the diagonal entries in $A$, i.e., $tr(A) = \sum_i A_{ii}$. Given a matrix $W = \{W_{ij}\}$ of size $m \times n$, the Frobenius norm of $W$ is the square root of the squared sum over all entries, i.e., $\|W\|_F = \sqrt{\sum_{ij}(W_{ij})^2}$. Following common notation, $W^T$ denotes the transposed matrix of $W$.

Singular value decomposition (SVD) applies to any real-valued matrix $W$. Specifically, the result of SVD on $W$ includes three matrices, $U$, $\Sigma$ and $V$, such that $W = U\Sigma V$. Here, $U$, $\Sigma$, and $V$ are of size $m \times s$, $s \times s$, and $s \times n$ respectively, where $m$ and $n$ are the number of rows and columns in $W$ respectively, and $s$ is a positive integer no larger than $\min\{m, n\}$. Moreover, $U$ and $V$ are row-wise and column-wise orthogonal matrices respectively. $\Sigma$ is a diagonal matrix, which contains non-negative real numbers on the diagonal and zeros in all the other entries. These diagonal entries, $\{\lambda_1, \lambda_2, \ldots, \lambda_s\}$, are called eigenvalues of the matrix $W$. The number of non-negative eigenvalues is called the rank of $W$, denoted as $rank(W)$.

When the rows and columns in the matrix $W$ are correlated, the rank of the matrix $W$ can be smaller than $m$ and $n$. In such cases, we say that $W$ is a low rank matrix. For example, when a group of records tend to appear together in a query, the workload matrix $W$ often exhibits strong column correlations. Similarly, when one query can be expressed as the linear combination of other queries, $W$ has strong row correlations. Both cases can be exploited to reduce the noise level necessary to satisfy differential privacy, as we showed in Section 1. Next we present the Low Rank Mechanism, a general solution to enforce differential privacy on a batch of linear queries, which utilizes the low rank property of the workload matrix to reduce noise.

## 4. WORKLOAD DECOMPOSITION

In this section, we propose a general workload matrix decomposition technique that minimizes the error for a batch of linear queries. Recall that the example in Figure 1 shows that instead of adding noise to the original data or query results (i.e., methods NOD and NOR), it is sometimes possible to construct another linear basis that leads to higher overall query accuracy. To build such a basis, we partition the workload matrix $W$ into the product of two components, $B = \{B_{ij}\}$ of size $m \times r$ and $L = \{L_{jk}\}$ of size $r \times n$, such that $W = BL$. Note that $r$ can be larger than the rank of the workload matrix $W$. Given the matrix decomposition, we design general mechanism for adding noise to $LD$ ($D$ is the dataset), and analyze the expected squared error. We first formally define the concepts of *query scale* and *query sensitivity*, for a given decomposition $W = BL$.

DEFINITION 1. *Query Scale*
*Given a workload decomposition $W = BL$, the scale of the decomposition, denoted by $\Phi(B, L)$, is the squared sum of the entries in $B$, i.e., $\Phi(B, L) = \sum_{i,j} B_{ij}^2$.*

DEFINITION 2. *Query Sensitivity*
*Given a workload decomposition $W = BL$, the sensitivity of the decomposition, denoted by $\Delta(B, L)$, is the maximal absolute sum of any column in $L$, i.e., $\Delta(B, L) = \max_j \sum_i |L_{ij}|$.*

Since $W = BL$, the linear query batch can be answered by calculating $Q(D) = WD = BLD$. Unlike solutions NOD and

1355

NOR, we inject noise into the intermediate result $LD$ to enforce differential privacy. Since $LD$ is another group of linear queries, we can apply NOR on $Q'(D) = LD$ with Eq. (5). The sensitivity of the new linear query batch is $\Delta(B, L)$, which leads to the following differential privacy mechanism $M_P(Q, D)$ with respect to the workload decomposition $W = BL$.

$$M_P(Q, D) = B \left( LD + Lap\left(\frac{\Delta(B, L)}{\epsilon}\right)^r \right) \quad (6)$$

The error analysis of $M_P(Q, D)$ is complicated as its adds noise at an intermediate step. The following lemma shows that the error is linear in the query scale, and quadratic in the query sensitivity.

LEMMA 1. *The expected squared error of $M_P(Q, D)$ with respect to the decomposition $W = BL$ is $2\Phi(B, L)(\Delta(B, L))^2/\epsilon^2$.*

Accordingly, we reduce the problem to finding the optimal workload decomposition $W = BL$ that minimizes $\Phi(B, L)(\Delta(B, L))^2$. However, this optimization problem is difficult to solve, since the objective function is the product of $\Phi(B, L)$ and $\Delta(B, L)$, and $\Delta(B, L)$ may not be derivable. To address this problem, we first prove an interesting property of the workload decomposition, which implies that the exact query sensitivity is actually not important.

LEMMA 2. *Given a workload decomposition $W = BL$ and a positive constant $\alpha$, we can always construct another decomposition $W = B'L'$ such that $B' = \alpha B$ and $L' = \alpha^{-1}L$, satisfying*

$$\Phi(B, L)(\Delta(B, L))^2 = \Phi(B', L')(\Delta(B', L'))^2$$

According to the above lemma, the balance between scale and sensitivity is not important, as we can always build another equivalent workload decomposition with arbitrary sensitivity. This motivates us to formulate a new optimization program, which focuses on minimizing the query scale while fixing the query sensitivity. The following theorem formalizes this claim.

THEOREM 1. *Given the workload $W$, $W = BL$ is the optimal workload decomposition to minimize expected squared error if $(B, L)$ is the optimal solution to the following program:*

$$\begin{aligned}
\text{Minimize: } &\text{tr}(B^T B)\\
\text{s.t. } &W = BL\\
&\forall j \sum_i |L_{ij}| \leq 1
\end{aligned} \quad (7)$$

In the optimization problem above, we are allowed to specify the number of columns in the matrix $B$, i.e. the rank $r$ of the matrix product $BL$. This enables us to generate matrices of significantly lower rank than the strategy matrix proposed in [16]. We thus use *Low Rank Mechanism* to denote the general query processing scheme in Eq. (6), using the optimal decomposition solution to Formula (7).

### 4.1 Optimality Analysis

In this subsection, we analyze the optimality of our optimization formulation. Specifically, we show that the utility of our proposed mechanism almost reaches the known utility lower bound for linear queries under differential privacy [14].

LEMMA 3. *Given a workload matrix $W$ of rank $r$ with eigenvalues $\{\lambda_1, \ldots, \lambda_r\}$, the expected squared error of $M_P(Q, D)$ w.r.t. the optimal decomposition $W = B^*L^*$ in low rank mechanism is bounded above by $\sum_{k=1}^{r} \lambda_k^2 r/\epsilon^2$.*

Using the geometric analysis technique under orthogonal projection [14], the following lemma reveals a lower bound on the squared error for linear queries.

LEMMA 4. *Given a workload matrix $W$ of rank $r$ with eigenvalues $\{\lambda_1, \ldots, \lambda_r\}$, the expected squared error of any $\epsilon$-differential privacy mechanism is at least*

$$\Omega\left(\left(\frac{2^r}{r!}\prod_{k=1}^{r}\lambda_k\right)^{2/r} r^3/\epsilon^2\right)$$

Assume that all the eigenvalues $\{\lambda_1, \lambda_2, \ldots, \lambda_r\}$ of workload $W$ are ordered in non-ascending order. We use $C = \lambda_1/\lambda_r$ to denote the ratio between the largest eigenvalue and the smallest non-zero eigenvalue. The following theorem discusses the tightness of low rank mechanism on error minimization. In particular, it proves the optimality of the result decomposition $W = B^*L^*$ with respect to Formula (7).

THEOREM 2. *When $r > 5$, the mechanism $M_p(Q, D)$ using $W = B^*L^*$ is an $O(C^2 r)$-approximately optimal solution w.r.t. the set of all non-interactive $\epsilon$-differential privacy mechanisms.*

When $C$ is close to 1, all non-zero eigenvalues are close to each other and the mechanism under our decomposition optimization program outputs results that well approximate the lower bound. This result answers one of the questions in [14], in which the authors discussed possible orthogonal projections but did not provide a concrete algorithm to identify the optimal projection. Our formulation can be regarded as an implementation of orthogonal projection with almost constant approximation. Therefore, our result fills the gap between theory and practice.

### 4.2 Relaxation on Decomposition

Theorem 2 shows that our decomposition leads to results with a tight bound. However, when there are very small eigenvalues in the workload matrix $W$, the bound in the theorem becomes loose. On the other hand, these small eigenvalues contribute little to the workload matrix $W$. This observation motivates us to design a new optimization formulation, in which $BL$ does not necessarily match $W$, but within a small error tolerance. This enables the formulation to find a more compact decomposition, such that the $r$ used in $B$ and $L$ can be smaller than the actual rank of $W$.

To do this, we introduce a new parameter $\gamma$ to bound the difference between $W$ and $BL$ in terms of the Frobenius norm. This leads to a new optimization problem:

$$\begin{aligned}
\text{Minimize: } &\text{tr}(B^T B)\\
\text{s.t. } &\|W - BL\|_F \leq \gamma\\
&\forall j \sum_i |L_{ij}| \leq 1
\end{aligned} \quad (8)$$

After finding the optimal $(B, L)$ for the problem in Formula 8, the mechanism $M_P(Q, D)$ outputs query results using Eq. (6). The error of this new mechanism is also bounded, as stated in the following theorem.

THEOREM 3. *The expected squared error of $M_P(Q, D)$ using the decomposition $(B, L)$ satisfying Eq. (8) is at most*

$$2\text{tr}(B^T B)/\epsilon^2 + \gamma \sum_i x_i^2$$

**Algorithm 1** Workload Matrix Decomposition

1: Initialize $\pi^{(0)} = \mathbf{0} \in \mathbb{R}^{m \times n}, \beta^{(0)} = 1, k = 1$
2: **while** not converged **do**
3: //Approximately solve the subproblem
4: **while** not converged **do**
5: $\quad B^{(k)} \leftarrow$ update $B$ using Eq. (9)
6: $\quad L^{(k)} \leftarrow$ run Algo. 2 to update $L$ w.r.t. Formula (10)
7: Compute $\tau = \|W - B^{(k)} L^{(k)}\|_F$
8: **if** $\tau$ is sufficiently small or $\beta$ is sufficiently large **then**
9: $\quad$ return $B^{(k)}$ and $L^{(k)}$
10: **if** $k$ is divisible by 10 **then**
11: $\quad \beta^{(k+1)} = 2\beta^{(k)}$
12: $\pi^{(k+1)} = \pi^{(k)} + \beta^{(k+1)} \left( W - B^{(k)} L^{(k)} \right)$
13: $k = k + 1$

**Algorithm 2** Nesterov's Projection Gradient Method

1: input: $\mathcal{G}(L), \frac{\partial \mathcal{G}}{\partial L}, L^{(0)}$
2: $\chi = r \cdot n \cdot 10^{-12}$, Lipschitz parameter: $\omega^{(0)} = 1$
3: Initializations: $L^{(1)} = L^{(0)}, \delta^{(-1)} = 0, \delta^{(0)} = 1, t = 1$
4: **while** not converged **do**
5: $\quad \alpha = \frac{\delta^{(t-2)} - 1}{\delta^{(t-1)}}, S = L^{(t)} + \alpha(L^{(t)} - L^{(t-1)})$
6: $\quad$ **for** $j = 0$ to ... **do**
7: $\qquad \omega = 2^j \omega^{(t-1)}, U = S - \frac{1}{\omega} \nabla_S$
8: $\qquad$ Project $U$ to the feasible set to obtain $L^{(t)}$ (i.e. solve Formula (11))
9: $\qquad$ **if** $\|S - L^{(t)}\|_F < \chi$ **then**
10: $\qquad\quad$ return;
11: $\qquad$ Define function: $\mathcal{J}_{\omega,S}(U) = \mathcal{G}(S) + \langle \frac{\partial \mathcal{G}}{\partial U}, U - S \rangle + \frac{\omega}{2} \|U - S\|_F^2$
12: $\qquad$ **if** $\mathcal{G}(L^{(t)}) \leq \mathcal{J}_{\omega,S}(U)$ **then**
13: $\qquad\quad \omega^{(t)} = \omega; L^{(t+1)} = L^{(t)}$; break;
14: $\quad$ Set $\delta^{(t)} = \frac{1 + \sqrt{1 + 4(\delta^{(t-1)})^2}}{2}$
15: $\quad t = t + 1$
16: return $L^{(t)}$

While Theorem 3 implies the possibility of estimating the optimal $\gamma$, it is not practical to implement it directly, because this estimation depends on the data, i.e., $\sum_i x_i^2$. In our experiments, we test different values of $\gamma$, and report their relative performance, regardless of the data distribution.

## 5. DECOMPOSITION ALGORITHM

The previous section formulates the workload matrix decomposition problem as an optimization program, which is rather complicated and non-trivial to solve. This section describes an effective and efficient solution for this program, based on the inexact Augmented Lagrangian Multiplier (ALM) method [5, 18].

The main challenge in solving the optimization program of Formula (8) is the non-smooth $\mathcal{L}_1$ regularized term. The projected gradient method [10] is considered one of the most efficient general algorithms to solve these problems. Following the strategy used in [5], we treat the $\mathcal{L}_1$ regularized term separately and approximately minimize a sequence of Lagrangian subproblems. Our inexact Augmented Lagrangian method for workload matrix decomposition problem is summarized in Algorithm 1.

In order to handle the linear constraints $\|W - BL\|_F \leq \gamma \to 0$, in which $W \in \mathbb{R}^{m \times n}$, $B \in \mathbb{R}^{m \times r}$ and $L \in \mathbb{R}^{r \times n}$, the inexact Augmented Lagrangian method introduces a positive penalty item $\beta \in \mathbb{R}$ and the Lagrange multiplier $\pi \in \mathbb{R}^{m \times n}$. The update on $\beta$ and $\pi$ follows the standard strategy used in [5, 18]. Given fixed $\beta$ and $\pi$ in each iteration, the algorithm aims to find a pair of new $B$ and $L$ to minimize the following subproblem:

$$\mathcal{J}(B, L, \beta, \pi) = \frac{1}{2} \text{tr}(B^T B) + \langle \pi, W - BL \rangle + \frac{\beta}{2} \|W - BL\|_F^2$$
$$\text{s.t. } \forall j \sum_i |L_{ij}| \leq 1$$

This is a Bi-Convex optimization problem, which can be solved by block gradient descent via alternately optimizing $B$ and $L$. Based on the formulation above, optimizing $B$ is straightforward. Since the gradient with respect to $B$ can be computed as:

$$\frac{\partial \mathcal{J}}{\partial B} = B - \pi L^T + \beta B L L^T - \beta W L^T$$

,
based on the fact that $\mathcal{J}(\cdot)$ is convex with respect to $B$, we can set $\frac{\partial \mathcal{J}}{\partial B} = 0$, and obtain a closed form solution to update $B$:

$$B = \left( \beta W L^T + \pi L^T \right) \left( \beta L L^T + I \right)^{-1} \quad (9)$$

The second step is to optimize $L$, which is equivalent to solving the following quadratic programming problem:

$$\mathcal{G}(L) = \frac{\beta}{2} \text{tr}\left( L^T B^T B L \right) - \text{tr}\left( (\beta W + \pi)^T B L \right)$$
$$\text{s.t. } \forall j \sum_i |L_{ij}| \leq 1 \quad (10)$$

In order to minimize Eq. (10) under constraints, we employ Nesterov's first order optimal method [23] to accelerate the gradient decent. Nesterov's method has a much faster convergence rate than traditional methods such as the subgradient method or the naïve projected gradient descent. In particular, the gradient of $\mathcal{G}(L)$ with respect to $L$ is

$$\frac{\partial \mathcal{G}}{\partial L} = \beta B^T B L - \beta B^T W - B^T \pi$$

$L$ is updated by gradient descent while ensuring that the $\mathcal{L}_1$ regularized constraint on $L$ is satisfied. This can be done by solving the following optimization problem:

$$\min_L \|L - L^{(t)}\|_F^2, \text{s.t. } \forall j \sum_i |L_{ij}| \leq 1, \quad (11)$$

in which $L^{(t)}$ denotes the last feasible solution after exactly $k$ iterations. Since Formula (11) can be decoupled into $r$ independent $\mathcal{L}_1$ regularized sub-problems, it can be solved efficiently by $\mathcal{L}_1$ projection methods [10]. The complete algorithm for the projection method is summarized in Algorithm 2.

**Convergence Analysis:** In each iteration, the algorithm solves a sequence of Lagrangian subproblems by optimizing $B$ (step 5) and $L$ (step 6) alternatively. The algorithm stops when a sufficiently small $\gamma$ is achieved or the penalty parameter $\beta$ is sufficiently large. It suffices to guarantee that $L$ converges to the optimal solution [18]. Although the objective function is non-smooth, the algorithm possesses excellent convergence properties. To be precise, we formally establish the following convergence statement.

THEOREM 4. *If $(B^{(k)}, L^{(k)})$ is the temporary solution after the $k$-th iteration and $(B^*, L^*)$ is the optimal solution to Formula (7), we have*

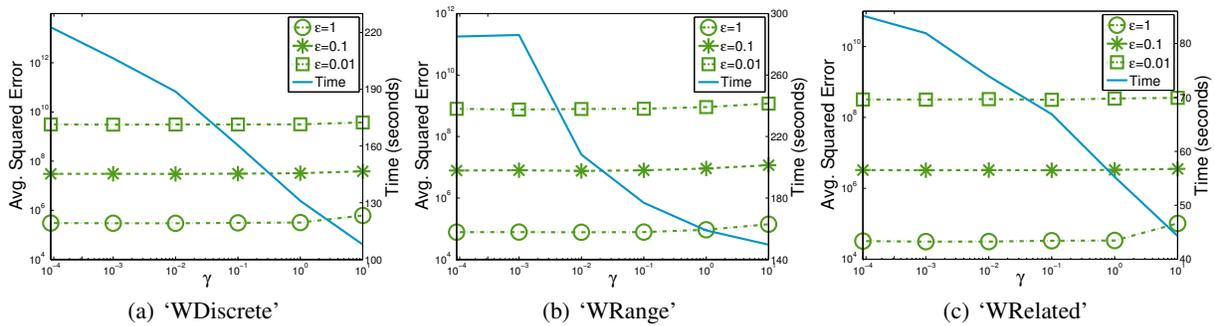

**Figure 2: Effect of varying relaxation parameter $\gamma$ with the *Search Logs* dataset for LRM**

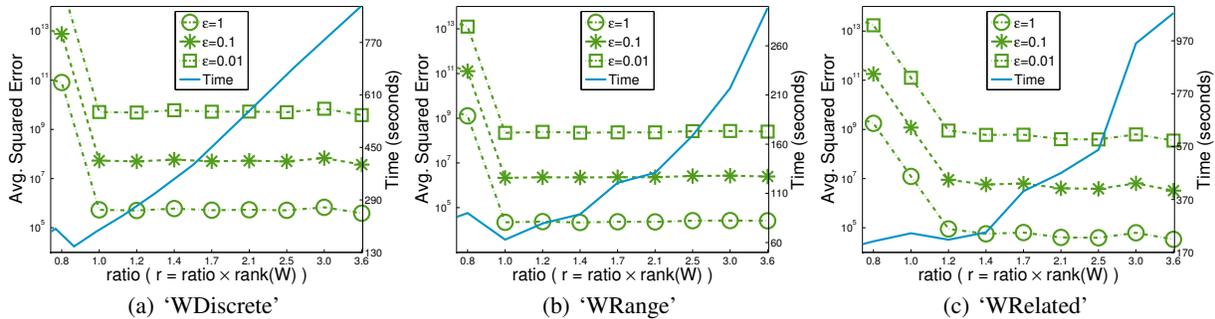

**Figure 3: Effect of varying $r$ with *Search Logs* dataset for LRM**

$$\left| tr(B^{(k)}B^{(k)}) - tr(B^*B^*) \right| \leq O\left(\frac{1}{\beta^{k-1}}\right) \quad (12)$$

Since $\beta^{(k)}$ doubles after every 10 iterations, the algorithm converges rapidly. This proves the fast convergence property of our algorithm.

**Complexity Analysis:** The total number of variables in $B$ and $L$ is $(m+n)r$. Each update on $B$ in Eq. (9) takes $O(r^2m)$ time, while each update on $L$ takes $O(r^2n)$ time. If Algorithm1 converges to a local minimum with $N_{in}$ inner iterations (at line 4 in Algorithm 1) and $N_{out}$ outer iterations (at line 2 in Algorithm 1), the total complexity of Algorithm 1 is $O(N_{in} \times N_{out} \times (r^2m + r^2n))$.

## 6. EXPERIMENTS

This section demonstrates the effectiveness of the proposed Low-Rank Mechanism (LRM), and compares it against four state-of-the-art methods: the Matrix Mechanism (MM) [16], the Laplace Mechanism (LM) [11], the Wavelet Mechanism (WM) [28] and the Hierarchical Mechanism (HM) [15]. We implemented the Matrix Mechanism (MM) by optimizing the $\mathcal{L}_2$ approximation instead of $\mathcal{L}_1$ error as suggested in [16]. The details of our MM implementation are available in Appendix B. All methods were implemented and tested in Matlab on a desktop PC with Intel quad-core 2.50 GHz CPU and 4GBytes RAM. In all experiments, every algorithm is executed 20 times and the average performance is reported. We employ three popular real datasets used in [15, 29]: *Search Log*, *Net Trace* and *Social Network*. *Search Log* includes search keyword statistics collected from *Google Trends* and *American Online* between 2004 and 2010. *Social Network* gives the number of users in a social network site with specific degrees in the social graph. *Net Trace* is a statistical database containing the number of TCP packets related to particular IP addresses, which is collected from a university intranet. *Search Logs*, *Net Trace* and *Social Network* contain $2^{16} = 65,536$, $2^{15} = 32,768$ and $11,342$ entries respectively. The reader is referred to [15] for more details of these datasets. We published our Matlab implementations of all algorithms used in the experiments, as well as sample datasets, online at http://yuanganzhao.weebly.com/.

To evaluate the impact of data domain cardinality on real datasets, we transform the original counts into a vector of fixed size $n$ (domain size), by merging consecutive counts in order. Given the number $m$ of linear queries in the batch, we generate three different types of workloads, namely *WDiscrete*, *WRange* and *WRelated*. In *WDiscrete*, for each weight $W_{ij}$ of query $q_i$ in the batch, we randomly select $W_{ij} = 1$ with probability 0.02 and set $W_{ij} = -1$ otherwise. In *WRange*, a batch of range queries on the domain are generated, by randomly picking up the starting location $a$ and ending location $b$ following a uniform distribution on the domain. Given the interval $(a, b)$, we set $W_{ij}$ of query $q_i$ in the batch to 1 for every $a \leq j \leq b$ and all other weights to 0. Finally, for *WRelated*, we generate $s$ (discussed later) independent base queries $A$ of size $s \times n$, by randomly assigning weights to the queries under a standard $(0, 1)$-normal distribution. Another group of correlation matrix $C$ of size $m \times s$ are generated similarly. The final workload $W$ of size $m \times n$ is the product of $C$ and $A$.

We test the impact of five parameters in our experiments: $\gamma$, $r$, $n$, $m$ and $s$. $\gamma$ is the relaxation factor defined in Formula (8). $r$ is the number of columns in $B$ (and also the number of rows in $L$). $n$ is the size of the domain and $m$ is the number of queries in the batch. Finally, $s$ is the number of rows of queries in the base $A$, which is only used in the generation of *WRelated*. The range of all these five parameters is summarized in Table 1. Unless otherwise specified, the default parameters in bold are used. Moreover, we test three different privacy budgets, $\epsilon = 1, 0.1$ and $0.01$. Note that the squared error incurred by all the methods is quadratic in $1/\epsilon$.

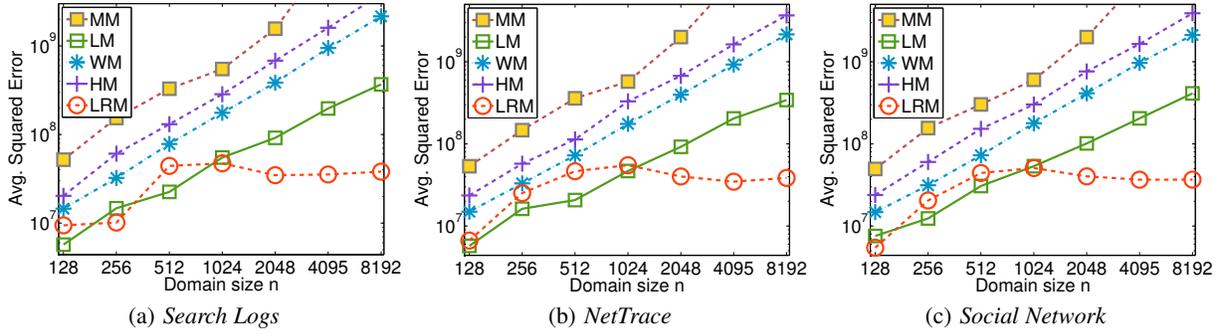

(a) *Search Logs*  (b) *NetTrace*  (c) *Social Network*

**Figure 4: Effect of varying domain size $n$ on workload 'WDiscrete' with $\epsilon = 0.1$**

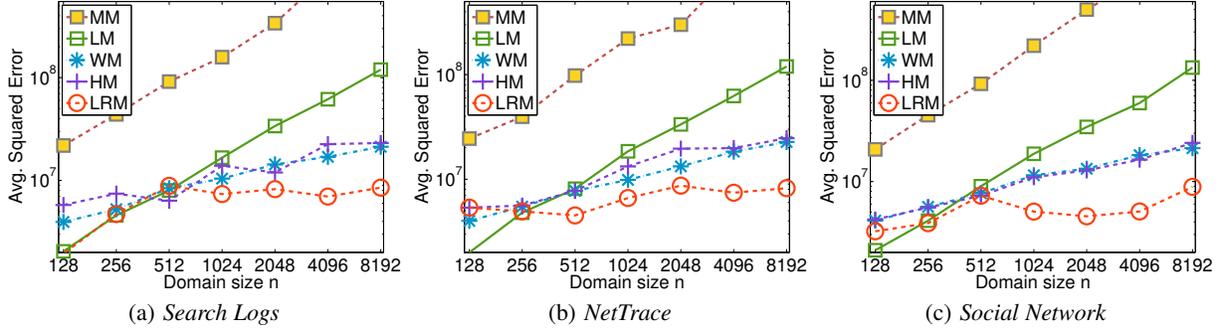

(a) *Search Logs*  (b) *NetTrace*  (c) *Social Network*

**Figure 5: Effect of domain size $n$ on workload *WRange* with $\epsilon = 0.1$**

| $\gamma$ | 0.0001, 0.001, **0.01**, 0.1, 1, 10 |
|---|---|
| $r$ | $\{0.8, 1.0, \mathbf{1.2}, 1.4, 1.7, 2.1, 2.5, 3.0, 3.6\} \times rank(W)$ |
| $n$ | 128, 256, 512, **1024**, 2048, 4096, 8192 |
| $m$ | 64, 128, **256**, 512, 1024 |
| $s$ | $\{0.1, 0.2, 0.3, 0.4, \mathbf{0.5}, 0.6, 0.7, 0.8, 0.9, 1.0\} \times min(m, n)$ |

**Table 1: Parameters used in the experiments**

In the experiments, we measure *Average Squared Error* and *Computation Time* of the methods. Specifically, the *Average Squared Error* is the average squared $\mathcal{L}_2$ distance between the exact query answers and the noisy answers. In the following, we first examine the impact of $\gamma$ and $r$, which are only used in the LRM method. The results provide important insights on how to tune these two parameters to maximize the utility of the LRM method.

### 6.1 Impact of $\gamma$ and $r$ on LRM

In LRM, $\gamma$ is an important parameter controlling the relaxation on the approximation of $BL$ to $W$. In our first set of experiments, we investigate the impact of $\gamma$ on the accuracy and the efficiency of LRM. Figure 2 reports the performance of LRM under all three different workloads, *WDiscrete*, *WRange* and *WRelated* on the *Search Log* dataset with varying values for $\gamma$. The results in the figure show that the errors of LRM on all three workloads are not sensitive to $\gamma$ in the range from $10^{-4}$ to 10. On the other hand, LRM executes much faster with larger $\gamma$. This suggests that a larger value for $\gamma$ is preferred in practice, to achieve high efficiency without losing much on result accuracy. Moreover, we also test with three different values of the privacy budget $\epsilon$. Since the decomposition method does not rely on $\epsilon$, the shapes of the result curves with different $\epsilon$ values are nearly identical, albeit at different scales. The average error is quadratic in the privacy budget $\frac{1}{\epsilon}$, as expected.

In LRM, $r$ is another important parameter that determines the rank of the matrix $BL$ that approximates the workload $W$. $r$ affects both the approximation accuracy and the optimization speed. When $r$ is too small, e.g., when $r < rank(W)$, our optimization formulation may fail to find a good approximation, leading to sub-optimal accuracy for the query batch. On the other hand, an overly large $r$ leads to poor efficiency, as the search space expands dramatically. We thus test LRM with varying $r$, by controlling the ratio of $r$ to the actual rank $rank(W)$, on the *Search Log* dataset. We record the average squared error under all the workloads and report it in Figure 3.

There are several important observations in Figure 3. First, when $r < rank(W)$, the accuracy of LRM is far worse (up to two orders of magnitude) than that in other settings. Second, the performance of LRM is rather stable when $r$ becomes larger than $1.2 \cdot rank(W)$. This is because the optimization formulation has enough freedom to find the optimal decomposition when $r$ is larger than $rank(W)$. Finally, the amount of computation spent on workload decomposition increases exponentially with $r$. Thus, to balance the efficiency and effectiveness of LRM, a good value for $r$ is between $rank(W)$ and $1.2 \cdot rank(W)$. We use the latter as the default value in the subsequent experiments.

### 6.2 Impact of Varying Domain Size $n$

We now evaluate the performance of all mechanisms with varying domain size $n$. As mentioned earlier in this section, the domain size is controlled by merging consecutive counts in the original domain. While different workloads and datasets are used, we only test with $\epsilon = 0.1$ because $\epsilon$ does not have much impact on the relative performance of different mechanisms. In Figures 4, 5 and 6, we report the result errors of all these mechanisms.

In all the experiments, the Matrix Mechanism (MM) is much worse than the other mechanisms, sometimes by an order of magnitude. This is mainly because 1) the strategy matrix in MM must be a full rank matrix; and 2) the $\mathcal{L}_2$ approximation used by MM does not lead to a good optimization of the actual objective function formulated using the error measure in $\mathcal{L}_1$. Because of its poor

1359

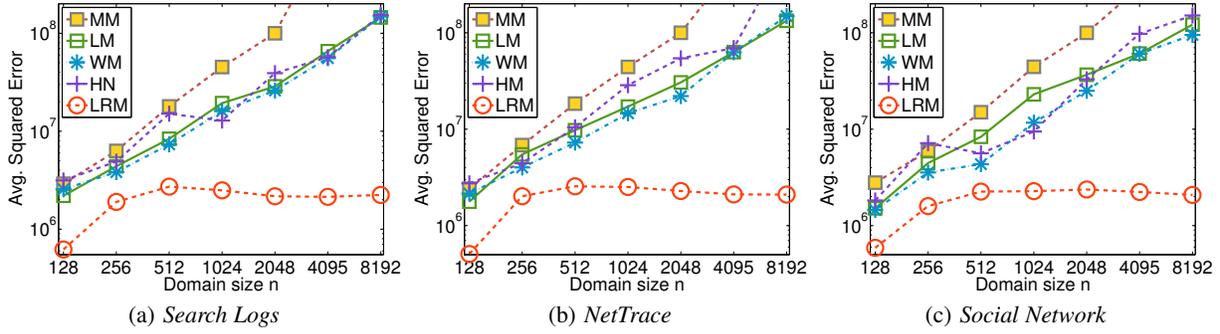

**Figure 6: Effect of domain size $n$ on workload *WRelated* with $\epsilon = 0.1$**

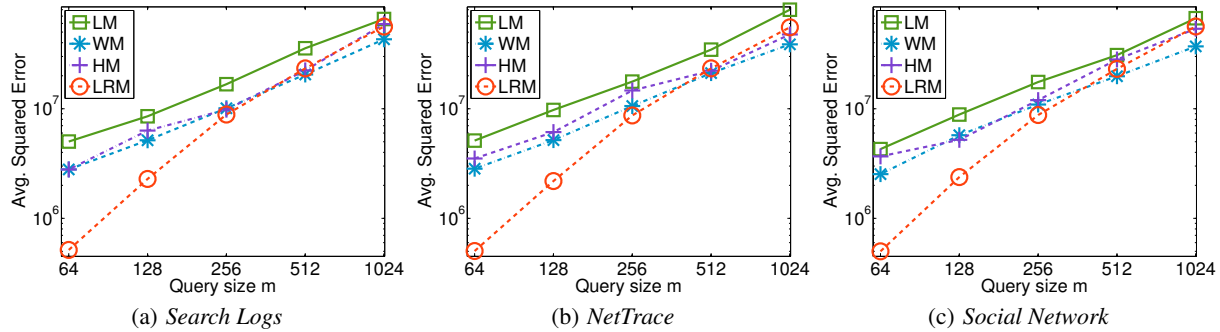

**Figure 7: Effect of number of queries $m$ on workload *WRange* with $\epsilon = 0.1$**

performance, we exclude MM in the rest of the experiments.

On the *WDiscrete* workload, the Laplace Mechanism (LM) outperforms all other mechanisms when the domain size is relatively small. This is in part due to the fact that the Wavelet Mechanism (WM) and the Hierarchical Mechanism (HM) are mainly designed to optimize range queries. While all other mechanisms incur linear error in terms of the domain size $n$, LRM's error stops increasing when the domain size is larger than 512. This is because LRM's error relies on the rank of the workload matrix $W$, and $rank(W)$ is no larger than $\min(m, n)$ no matter how large $n$ is. This explains the excellent performance of LRM on larger domains. On the *WRange* workload, the errors of WM and HM are smaller than LM when the domain size is no smaller than 512, in which case their strategies work better. LRM's performance is still significantly better than any of them, since LRM fully utilizes the correlations between these range queries on large domains. Finally, on the *WRelated* workload, LRM achieves the best performance on all test cases. The performance gap between LRM and other methods is over two orders of magnitude, when the domain size reaches 8192. Since *WRelated* naturally leads to a low rank workload matrix $W$, this result verifies LRM's vast benefit from exploiting the low-rank property of the workload.

### 6.3 Impact of Varying Query Size $m$

In this subsection, we test the impact of the query set cardinality $m$ on the performance of the mechanisms. We mainly focus on settings when the number of queries $m$ is no larger than the domain size $n$, i.e. $m \leq n$. Due to space limitations, we only present the results on *WRange* and *WRelated* workloads in Figures 7 and 8.

The results lead to several interesting observations. On *WRange* workload (Figure 7), LRM outperforms the other mechanisms, when the number of queries $m$ is significantly smaller than $n$. With growing $m$, the performance of all mechanisms on *WRange* tends to converge. When $m = 1024$, WM achieves the best performance among all mechanisms, since it is optimized for range queries. The degeneration in performance of LRM is due to the lack of low rank property when the batch contains too many random range queries. On *WRelated* workload, LRM is dramatically better than the other methods, for any query set cardinality $m$. Regardless of the value of $m$, the rank of the *WRelated* workload $W$ remains low, which is solely determined by the parameter $s$ used in the workload generation procedure. These results further confirm that the squared error generated by LRM scales linearly with the rank of the workload.

### 6.4 Impact of Varying Query Rank $s$

All previous experiments demonstrate LRM's substantial performance advantage when the workload matrix has low rank. In this group of experiments, we manually control the rank of workload $W$ to verify the correctness of our claim. Recall that the parameter $s$ determines the size of the matrix $C_{m \times s}$ and the size of the matrix $A_{s \times n}$ in the generation of the *WRelated* workload. When $C$ and $A$ contain only independent rows/columns, $s$ is exactly the rank of the workload matrix $W = CA$. In Figure 9, we vary $s$ from $0.1 \min(m, n)$ to $\min(m, n)$. Compared to the other mechanisms, LRM maintains an accuracy advantage of over two orders of magnitude, when the rank of the workload matrix is low. With increasing rank of $W$, the accuracy of other mechanisms remain stable, while LRM's error grows rapidly. This phenomenon again confirms that the low rank property is the main reason behind LRM's advantages with respect to error minimization.

## 7. CONCLUSION

This paper presented the *Low Rank Mechanism* (LRM), an optimization framework that minimizes the overall error in the results of a batch of linear queries under $\epsilon$-differential privacy. LRM is the first practical method for a large number of linear queries, with an efficient and effective implementation using well established optimization techniques. Experiments show that LRM significantly outperforms other state-of-the-art differentially private query processing mechanisms, often by orders of magnitude. The current design of LRM focuses on exploiting the correlations between different queries. One interesting direction for future work is to further optimize LRM by utilizing also the correlations between data values, e.g., as is done in [29, 24, 17].

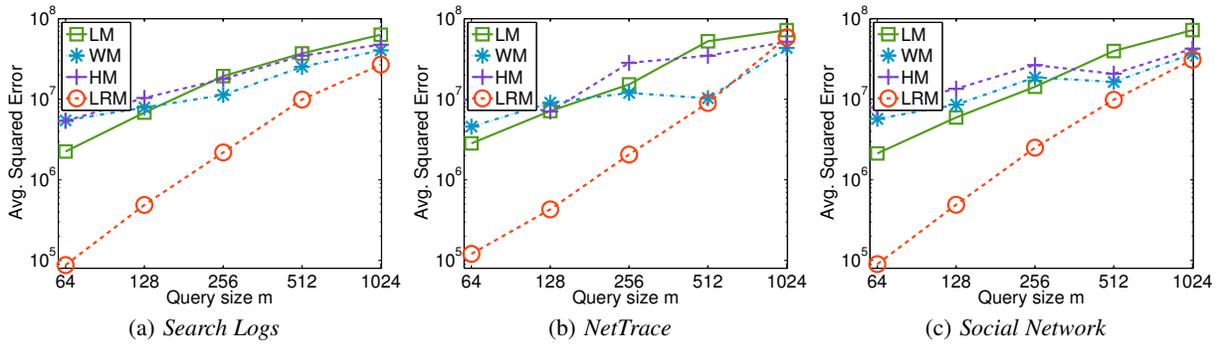

Figure 8: Effect of number of queries $m$ on workload *WRelated* with $\epsilon = 0.1$

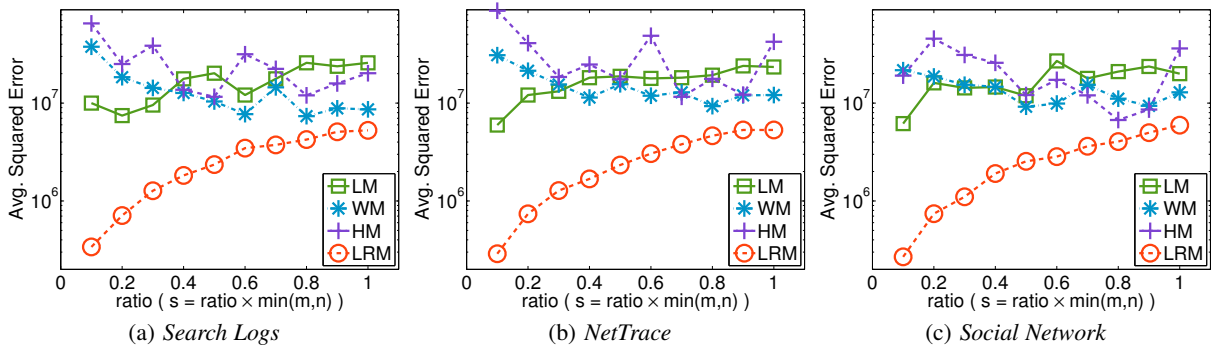

Figure 9: Effect of parameter $s$ with $\epsilon = 0.1$

## Acknowledgments

Yuan and Hao are supported by NSF-China (61070033, 61100148), NSF-Guangdong (9251009001000005, S2011040004804) and Key Technology Research and Development Programs of Guangdong Province (2010B050400011). Zhang, Winslett, Xiao and Yang are supported by SERC 102-158-0074 from Singapore's A*STAR. Xiao is also supported by SUG Grant M58020016 and AcRF Tier 1 Grant RG 35/09 from Nanyang Technological University.

# APPENDIX
## A. PROOFS

*Lemma 1:*

PROOF. Based on the definition of the mechanism in Eq. (6), the residual of the noisy result with respect to the exact result, i.e. $Q(D) - M_P(Q, D)$, is $B \cdot Lap\left(\frac{\Delta(B,L)}{\epsilon}\right)^r$. The expected squared error is thus $\sum_{ij} B_{ij}^2 \frac{2(\Delta(B,L))^2}{\epsilon^2}$. Since $\Phi(B, L) = \sum_{ij} B_{ij}^2$, the expected error of the mechanism is $2\phi(B, L)(\Delta(B, L))^2/\epsilon^2$. □

*Lemma 2:*

PROOF. Based on the definition of sensitivity, we have $\Delta(B', L') = \max_j \sum_i |L'_{ij}| = \max_j \sum_i |L_{ij}/\alpha| = \alpha^{-1}\Delta(B, L)$.

The last equality holds because $\alpha$ is a positive constant. On the other hand, the scales of the decompositions follow a similar relationship:

$$\Phi(B', L') = \sum_{ij}(B'_{ij})^2 = \sum_{ij} \alpha^2 (B_{ij})^2 = \alpha^2 \Phi(B, L)$$

Therefore, $\Phi(B', L')(\Delta(B', L'))^2 = \Phi(B, L)(\Delta(B, L))^2$. Finally, since $B'L' = BL = W$, we reach the conclusion of the lemma. □

*Theorem 1:*

PROOF. Assume that $(B^*, L^*)$ is the best matrix decomposition for minimizing the expected squared error for $M_P(Q, D)$. In the following, we prove that $(B^*, L^*)$ is optimal, if and only if it also minimizes the program in Formula (7).

(*if* part): If $(B, L)$ minimizes Formula (7) but $(B, L)$ incurs more expected error than $(B^*, L^*)$, implying that

$$\Phi(B^*, L^*)(\Delta(B^*, L^*))^2 < \Phi(B, L)(\Delta(B, L))^2$$

By applying Lemma 2, we can construct another decomposition $B' = \Delta(B^*, L^*)B^*$ and $L' = \Delta(B^*, L^*)^{-1}L^*$, such that $\Phi(B', L')(\Delta(B', L'))^2 < \Phi(B, L)(\Delta(B, L))^2$. On the other hand, since $\Delta(B', L') \leq 1$, we have $\max_j \sum_i |L'_{ij}| = 1$. Therefore, we can derive the following inequalities.

$$\begin{aligned} \Phi(B', L') &= \Phi(B', L')(\Delta(B', L'))^2 \\ &< \Phi(B, L)(\Delta(B, L))^2 \\ &\leq \Phi(B, L) \end{aligned}$$

Finally, since $\Phi(B', L') = \text{tr}(B'^T B')$ and $\Phi(B, L) = \text{tr}(B^T B)$, it leads to a contradiction if $\text{tr}(B'^T B') < \text{tr}(B^T B)$.

(*only if* part): If $(B^*, L^*)$ is not the optimal solution to the program in Formula (7), the optimal solution $(B, L)$ must incur less expected error, using a similar strategy. This completes the proof of the theorem. □

*Lemma 3:*

PROOF. To prove the lemma, we aim to artificially construct a workload decomposition $W = BL$ satisfying the constraints of the optimization formulation. If the error of this artificial decomposition is no larger than the upper bound, the exact optimal solution must render results with less error.

Recall that $W$ has a unique SVD decomposition $W = U\Sigma V$ such that $\Sigma$ is a diagonal matrix of size $r \times r$. We thus build a decomposition $B = \sqrt{r}U\Sigma$ and $L = \frac{1}{\sqrt{r}}V$, in which $r$ is the rank of the matrix $W$. First, we will show such $(B, L)$ satisfies the constraints in Formula (7). It is straightforward to show it satisfies the first constraint: $BL = \sqrt{r}U\Sigma\frac{1}{\sqrt{r}}V = U\Sigma V = W$.

Regarding the second constraint, since $V$ only contains orthogonal vectors, every column $j$ must have $\|V_{:j}\|_2 = \|v\|_2 = 1$. By the norm triangle inequality, $\|v\|_2 \leq \|v\|_1 \leq \sqrt{r}\|v\|_2$, and we obtain $\frac{1}{\sqrt{r}}\sum_i |V_{ij}| \leq 1$. Therefore, such $(B, L)$ must be a valid solution to the program.

The expected squared error of the artificial decomposition $W = BL$ is at most

$$\begin{aligned} \text{tr}(B^T B)/\epsilon^2 &= \text{tr}((\sqrt{r}U\Sigma)^T(\sqrt{r}U\Sigma))/\epsilon^2 \\ &= \text{tr}(\Sigma^T U^T U\Sigma))r/\epsilon^2 \\ &= \sum_{k=1}^{r} \lambda_k^2 r/\epsilon^2 \end{aligned}$$

This proves that $\sum_{k=1}^{r} \lambda_k^2 r/\epsilon^2$ is an upper bound for the noise of our decomposition-based scheme. □

*Lemma 4:*

PROOF. In Corollary 3.4 in [14], Hardt and Talwar proved that any $\epsilon$-differential privacy mechanism incurs expected squared error no less than[1] $\Omega(r^3 (Vol(PWB_1^n))^{2/r}/\epsilon^2)$.

In the formula above, $B_1^n$ is the $\mathcal{L}_1$-unit ball. $Vol(PWB_1^n)$ is the volume of the unit ball after the linear transformation under $PW$, in which $P$ is any orthogonal linear transformation matrix from $\mathbb{R}^m \mapsto \mathbb{R}^r$. To prove the lemma, we construct an orthogonal transformation $P$ using the SVD decomposition over $W = U\Sigma V$. By simply letting $P = U^T$, since $U^TU$ and $VV^T$ are identity matrices, we have $Vol(PWB_1^n) = Vol(PUVV^T\Sigma VB_1^n) = Vol(V(V^T\Sigma V)B_1^n) = Vol(VB_1^n)\prod_{k=1}^{r}\lambda_k$. The last equality holds due to Lemma 7.5 in [14]. Consider the the convex body $VB_1^n$. It is an $r$-dimensional unit ball after the orthogonal transformation under $V$. Note that $Vol(B_1^r)$ can be computed using the well known $\Gamma$ function, as in [26], $2^r\frac{\Gamma(2)}{\Gamma(1+r)} = \frac{2^r}{r!}$. Therefore, the lower bound can be computed as: $\Omega((\frac{2^r}{r!}\prod_{k=1}^{r}\lambda_k)^{2/r}r^3/\epsilon^2)$. This reaches the conclusion of the lemma. □

*Theorem 2:*

PROOF. To prove the theorem, we investigate the ratio of the upper bound to the lower bound.

---

[1] [14] used absolute error in the paper, which we change to squared error here.

$$\frac{\sum_{k=1}^{r} \lambda_k^2 r/\epsilon^2}{\left(\frac{2r}{r!} \prod_{k=1}^{r} \lambda_k\right)^{2/r} r^3/\epsilon^2}$$

$$\leq \frac{\sum_{k=1}^{r} \lambda_1^2}{\left(\frac{2r}{r!} \prod_{k=1}^{r} \lambda_r\right)^{2/r} r^2}$$

$$\leq \frac{r\lambda_1^2}{\left(\frac{2r}{r!}\right)^{2/r} \lambda_r^2 r^2} = \frac{C^2}{\left(\frac{2r}{r!}\right)^{2/r} r} \leq \left(\frac{C}{4}\right)^2 r$$

The last inequality holds due to the fact that $r! < \left(\frac{r}{2}\right)^r$ when $r > 5$. Note that all the inequalities above are tight, and the equalities hold when $C = 1$, i.e. $\lambda_1 = \lambda_2 = \ldots = \lambda_r$. Thus, we prove that the approximation factor of our decomposition scheme is $O(C^2 r)$. □

*Theorem 3:*

PROOF. When $W \neq BL$, the error has two parts. The first part is the noises due to the Laplace random variables. Using Lemma 1, the incurred error is at most $\frac{2}{\epsilon^2}\Phi(B,L)(\Delta(B,L))^2 \leq \frac{2}{\epsilon^2}\text{tr}(B^T B)$.

The second part of the error is the structural error on the results. The expected squared error is measured as

$$((W - BL)D)^T(W - BL)D$$

$$\leq \|W - BL\|_F^2 D^T D = \|W - BL\|_F^2 \sum_{i=1}^{n} x_i^2$$

The inequality is due to the Cauchy Schwartz inequality. By linearity of expectation, the expected squared errors can be simply summed up. This leads to the conclusion of the theorem. □

*Theorem 4:*

PROOF. We use $B^{(k)*}$ to denote the optimal solution of the Lagrangian sub-problem in $k^{th}$ iteration. Note the following inequality on the sequence of the Lagrangian subproblems:

$$\mathcal{J}(B^{(k+1)*}, L^{(k+1)*}, \pi^{(k)*}, \beta^{(k)})$$
$$= \min_{B,L} \mathcal{J}(B, L, \pi^{(k)*}, \beta^{(k)})$$
$$\leq \min_{\|W-BL\|_F \leq \gamma, \forall j \sum_i |L_{ij}| \leq 1} \mathcal{J}(B, L, \pi^{(k)*}, \beta^{(k)})$$
$$= \min_{\|W-BL\|_F \leq \gamma, \forall j \sum_i |L_{ij}| \leq 1} \frac{1}{2}\text{tr}(B^T B) = \frac{1}{2}\text{tr}(B^{*T} B^*)$$

Based on the above inequality, we derive the following inequality:

$$\frac{1}{2}\text{tr}(B^{(k+1)T} B^{(k+1)})$$
$$= \mathcal{J}(B^{(k+1)*}, L^{(k+1)*}, \pi^{(k)*}, \beta^{(k)}) - \langle \pi^{(k)}, W - B^{(k+1)} L^{(k+1)} \rangle + \frac{\beta^{(k)}}{2}\|W - B^{(k+1)} L^{(k+1)}\|_F^2$$
$$= \mathcal{J}(B^{(k+1)*}, L^{(k+1)*}, \pi^{(k)*}, \beta^{(k)}) - \frac{1}{2\beta^{(k)}}(\|\pi^{(k)} + \beta^{(k)}(W - B^{(k+1)} L^{(k+1)})\|_F^2 - \|\pi^{(k)}\|_F^2)$$
$$= \mathcal{J}(B^{(k+1)*}, L^{(k+1)*}, \pi^{(k)*}, \beta^{(k)}) - \frac{1}{2\beta^{(k)}}(\|\pi^{(k+1)}\|_F^2 - \|\pi^{(k)}\|_F^2)$$
$$\leq \frac{1}{2}\text{tr}(B^{*T} B^*) - \frac{1}{2\beta^{(k)}}\left(\|\pi^{(k+1)*}\|_F^2 - \|\pi^{(k)*}\|_F^2\right)$$

The third equality holds because of the Lagrangian multiplier update rule:

$$W - B^{(k+1)*} L^{(k+1)*} = \frac{1}{\beta^{(k)}}\left(\pi^{(k+1)*} - \pi^{(k)*}\right)$$

Since $\pi^{(k)*}$ is always bounded, we conclude that

$$\frac{1}{2}\text{tr}\left(B^{(k+1)T} B^{(k+1)}\right) - \frac{1}{2}\text{tr}\left(B^{*T} B^*\right) \leq O\left(\frac{1}{\beta^{(k)}}\right)$$

This completes the proof of the theorem. □

## B. IMPLEMENTATION OF THE MATRIX MECHANISM

In [16], Li et al propose the *Matrix Mechanism*. The core of their method is finding a matrix $A$ to minimize the following the program.

$$\min_{A \in R^{r \times n}} \|A\|_2^2 \text{tr}(W^T W A^\dagger A^{\dagger T}) \tag{13}$$

Li et. al. [16] present a complicated implementation that is rather impractical due to its prohibitively high complexity. We hereby present a simpler and more efficient solution to their optimization program. Here $\|A\|_2^2$ denotes the maximum $\mathcal{L}_2$ norm of column vectors of $A$, therefore $\|A\|_2^2 = \max(\text{diag}(A^T A))$. Since $(A^T A)^{-1} = (A^T A)^\dagger$ ($A$ has full column rank), we let $M = A^T A$, and reformulate Formula (13) as the following semidefinite programming problem:

$$\min_{M \in R^{n \times n}} \max(\text{diag}(M))\text{tr}(W^T W M^{-1}) \; s.t. \; M \succ 0$$

$A$ is given by $A = \sum_i^n \sqrt{\lambda_i} v_i v_i^T$, where $\lambda_i, v_i$ are the $i$th eigenvalue and eigenvector of $M$, respectively. Calculating the second term $\text{tr}(W^T W M^{-1})$ is relatively straightforward. Since it is smooth, its gradient can be computed as $-M^{-1} W^T W M^{-1}$. However, calculating the first term $\max(\text{diag}(M))$ is harder since it is non-smooth. Fortunately, inspired by [7], we can still use a logarithmic and exponential function to approximate this term.

**Approximate the maximum positive number:** Since $M$ is positive definite, $v = \text{diag}(M) > 0$. we let $\mu > 0$ and define:

$$f_\mu(v) = \mu \log \sum_i^n \left(\exp\left(\frac{v_i}{\mu}\right)\right) \tag{14}$$

We then have $\max(v) \leq f_\mu(v) \leq \max(v) + \mu \log n$. If we set $\mu = \frac{\epsilon}{\log n}$, this becomes a uniform $\epsilon$-approximation of $\max(v)$ with a Lipschitz continuous gradient with constant $\omega = \frac{1}{\mu} = \frac{\log n}{\epsilon}$. The gradient of the objective function with respect to $v$ can be computed as:

$$\frac{\partial f}{\partial v_i} = \frac{\exp\left(\frac{v_i - \max(v)}{\mu}\right)}{\sum_j^n \left(\exp\left(\frac{v_j - \max(v)}{\mu}\right)\right)} \tag{15}$$

To mitigate the problems with large numbers, using the property of the logarithmic and exponential functions, we can rewrite Eq. (14) and Eq. (15) as:

$$f_\mu(v) = \max(v) + \mu \log \left(\sum_i^n \exp\left(\frac{v_i - \max(v)}{\mu}\right)\right)$$

$$\frac{\partial f}{\partial v_i} = \left(\sum_j^n \exp\left(\frac{v_j - v_i}{\mu}\right)\right)^{-1}$$

This formulation allows us to run the non-monotone projected gradient descent algorithm [2] and iteratively improves the result.